\documentclass[12pt]{article}
\usepackage{amsmath,amssymb,pgf,pgfarrows,pgfnodes,subfig,float,appendix, hyperref}
\usepackage[margin=0.7in]{geometry}

\title{{\rm\footnotesize \qquad \qquad \qquad \qquad \qquad \ \qquad \qquad \qquad \ \ \ \ \ \                  UTTG-15-12\ TCC-015-12     RUNHETC-2012-17     
SCIPP 12/11}\vskip.5in     Holographic Space-time Does Not Predict Firewalls}
\author{Tom Banks\\
Department of Physics and SCIPP\\
University of California, Santa Cruz, CA 95064\\
{\it and}\\
Department of Physics and NHETC\\
Rutgers University, Piscataway, NJ 08854\\
E-mail: \href{mailto:banks@scipp.ucsc.edu}{banks@scipp.ucsc.edu}
\\
\\
Willy Fischler\\
Department of Physics and Texas Cosmology Center\\
University of Texas, Austin, TX 78712\\
E-mail: \href{mailto:fischler@physics.utexas.edu}{fischler@physics.utexas.edu}}

\begin{document}
\maketitle

\begin{abstract}
We use the formalism of Holographic Space-time (HST) to investigate the claim of \cite{amps} that old black holes contain a firewall, {\it i.e.} an in-falling detector encounters highly excited states at a time much shorter than the light crossing time of the Schwarzschild radius\footnote{In 1996, N. Itzhaki\cite{itzhaki} claimed that quantum considerations led to singular black hole horizons.  His arguments were not related to those of \cite{amps} and made no distinction between young and old black holes.  Our discussion does not apply to Itzhaki's argument.}.  This conclusion is much less dramatic in HST than in the hypothetical models of quantum gravity used in \cite{amps}. In HST there is no dramatic change in particle physics inside the horizon until a time of order the Schwarzschild radius.\end{abstract}

\section{Introduction}

A recent paper \cite{amps} has made a dramatic new claim about black hole physics, and has led to a burgeoning literature\cite{boussoharlow}.  The classical physics of black holes indicates that an in-falling observer will encounter a singularity in a time of order the light crossing time of the Schwarzschild radius, $R_S$.  While the geometry near the singularity is unstable to small perturbations, the geometry a Schwarzschild radius away is not.  It has therefore been assumed that this conclusion holds for black holes that have evolved from larger holes by Hawking evaporation.

Using some general ideas about what a model of quantum gravity should look like, the authors of \cite{amps} (AMPS) claim that for an old black hole the claim of the previous paragraph is not true.  The adjective old means that the hole has evaporated for a time of order its Page time\cite{page}, the time over which the information in the black hole is transmitted to the Hawking radiation.  The Page time scales like the cube of $R_S$, which is of order the lifetime of the black hole.  They claim instead that an observer which falls into the black hole after the Page time encounters a ``firewall", a region in which the local observer is subjected to a barrage of high energy radiation, shortly after it crosses the horizon.  This conclusion has been seconded\cite{ampsrefs}, but most authors disagree\cite{boussoharlow}.
However, many of those authors do argue that something
highly non-classical happens near the horizon of an old black hole.

In fact, there is only one extant theory of QG, which discusses the observations of local detectors\cite{HST}.  The analyses of \cite{amps} \cite{ampsrefs} and \cite{boussoharlow} assume a theory with a single Hilbert space shared by all the detectors, which has approximate local observables in it.  They discuss the entanglement of different subsystems in this Hilbert space.
HST does not even allow us to define that entanglement.

Instead, HST describes space-time in terms of {\it an infinite number of independent quantum systems, each of which describes the proper time evolution along a particular time-like trajectory.}  The dynamics along each trajectory defines the quantum notion of a {\it causal diamond} because it couples together only a finite number of degrees of freedom during a finite amount of proper time.  Space-time as a whole is knit together by prescribing overlaps between the causal diamonds along one trajectory and those along others.  These overlaps are tensor factors in the Hilbert spaces describing individual diamonds.  The fundamental dynamical principle of HST is that {\it for each overlap, the density matrix defined by the dynamics along one trajectory, is unitarily equivalent to the density matrix defined by the other trajectory's dynamics.}  

Despite this, it turns out to be possible to discuss the AMPS firewall in terms of the experience of a single observer, which orbits the black hole until it has shrunk to less than half its area, and then falls in.   The resolution of the paradox in this observer's Hilbert space has to do with the fundamental difference between HST and the hypothetical quasi-field theoretical model of QG assumed in \cite{amps}. This is the distinction between particle degrees of freedom and the rest of the variables of QG.  In \cite{amps}, DOF that are not well described by QFT are assumed to be ``high energy".  In contrast, as we will describe in the next section, HST assigns $o(N^2 )$ DOF to a causal diamond whose holographic screen has an area of $o(N^2 )$ in Planck units.  Only $o(N^{3/2} )$ of these DOF have a QFT-like character.  The others, which we call {\it horizon DOF}, actually have low energy in the Hamiltonian that describes a geodesic observer in flat space, but are decoupled from the particles as $N\rightarrow\infty$.  {\it The distinction between particle and horizon degrees of freedom depends on the observer.  Particles are always defined as states that satisfy a certain constraint.  The Hamiltonians of different observers preserve that constraint to a greater or lesser degree.}  In \cite{holounruh} we will show that this gives an explanation of the Unruh effect.  In the present context what is important is that {\it any} state in a causal diamond that can be described in terms of particle physics, even in Minkowski space, is a state of low entropy, but that the entropy deficit can never be larger than $N^{3/2}$.  The semiclassical description of a black hole interior implies that it can remain in such a low entropy state for a time of order its Schwarzschild radius.  The state of a young black hole is therefore non-generic, yet it can still appear to be a completely thermalized object from the outside, because $N^2 - N^{3/2} \sim N^2$ for large $N$.  We will show in this paper that none of the overlap conditions of HST force us to deviate from the contention that the old black hole is in a non-generic state, where particle physics remains valid for a time of order the Schwarzschild time.   AMPS assumed that the young black hole was in a generic state, so that a measurement of more than half its information content completely determined the state of the whole system.  In HST, this assumption is incompatible with the claim that the young black hole interior was describable by particle physics for a Schwarzschild time.

The basic idea of HST is to construct quantum systems compatible with the space-time structure of a given classical space-time.  The conventional picture of an evaporating black hole is that of a Schwarzschild metric with a decreasing mass.  This includes the proviso that time-like trajectories inside the evolving horizon do not encounter a singularity for a finite time.  A causal diamond whose past tip is near the horizon and whose future tip is a proper time of order $R_S$ later, should have a description close to that of a similar diamond in Minkowski space, if $R_S$ is large.   As a consequence, the description of this black hole, as seen even by an external observer {\it must} find the black hole in a non-generic state, in which the particles inside the black hole can remain decoupled from the horizon degrees of freedom, at least for a time of order $R_S$.   This continues to be true as the black hole evaporates.   It follows from the fact that the causal diamond of the external observer is space-like separated from causal diamonds inside the horizon of the black hole, plus the implicit assumption of the space-time picture of the interior, that a particle description of the interior is valid.
As we will see, in HST this is equivalent to requiring that the state in the external observer's Hilbert space is a tensor product of a particle state in the interior, and a state in the combined Hilbert space of asymptotic particles and horizon states.

It is extremely important that the entropy of the interior particle Hilbert space scales like $R_S^{3/2} \ll R_S^2$, the total entropy of the black hole.  This implies that the non-generic nature of the black hole state does not change its thermodynamic behavior.
Eventually, when $R_S \sim 1$, the separation of black hole states into interior particles and horizon states ceases to make sense.  At this point the state is generic and all the entropy can be read from the outgoing radiation, which is entangled with it.
It's also clear that the decoupling between particles inside the horizon, and horizon states, cannot persist indefinitely, if the black hole is to evaporate.  Fortunately, but not coincidentally, the space-time geometry of the black hole gives us the same message.
In a time of order $R_S$, any interior geodesic hits the singularity, an event which we interpret as a change of the time dependent Hamiltonian along that trajectory, to one in which particle states mix strongly with the horizon and thermalize with it.   In this way, the descriptions of infalling and orbiting observers become consistent with each other, even as the black hole evaporates.   We will describe more details of this picture, and of the constraints it puts on the use of local QFT, in later sections.

The remaining worry is that if one has waited a Page time in the exterior before falling into the black hole, then the black hole really was in a generic state, because the Page time is so much longer than the infall time.  Here is where the relativity of simultaneity becomes really important.  HST implements the relativity of simultaneity by assigning a different Hilbert space and time-dependent Hamiltonian to each time-like trajectory.  The different systems are coupled by the overlap conditions, so the proper statement of the remaining worry is that the overlap conditions might force us to conclude that the late falling observer could only have a description compatible with that of the early falling observer, if there were a firewall.  We will show that this is not true, because the overlaps of the causal diamonds of the early and late falling observers are small, until the late observer has traveled of order the Schwarzschild radius.

In the next section, we will sketch the formalism of HST, providing only enough detail to appreciate our argument in the heart of the paper.  Our conclusion is simply stated: within the HST formalism we find no evidence for the existence of a firewall.   For an observer falling into a newly formed black hole, the particle degrees of freedom begin to interact with the horizon DOF in a time of order $R_S$.  After this time, the Hamiltonian for this observer begins to vary on the Planck time scale, and time can no longer be identified with proper time along a geometrical trajectory.  The notion of geometry has broken down. The same is true for an old black hole.    As time goes on, more and more DOF become rapidly time dependent, and in a time of order $R_S$ this time dependence also affects the ``particles".  In fact, after a time of order $R_S$, the distinction between particles and horizon DOF is no longer possible.

Throughout this paper, all distances and times will be expressed in Planck units. Events do not occur in real time...

\section{HST, Sketchily}

This section is intended to adumbrate the formalism of HST, which has been extensively reviewed in\cite{HST}.  The fundamental element of the theory is the quantum mechanics of a single time-like trajectory in space-time.  This is a (proper) time dependent unitary evolution operator $U(t_{f\ n} , t_{i\ n})$ which incorporates the requirements of causality.  $U$ operates in a Hilbert space associated with the beginning and end of the maximal causal diamond of that trajectory.  At intermediate times, $U$ factorizes into an operator acting on a small tensor factor of this Hilbert space, associated with the smaller causal diamond of this interval of proper time, and one operating on its tensor complement.  The dimensions of these Hilbert spaces, when they are large, determine the areas of the holographic screens of the associated diamonds, according to the BHFSB\cite{BHFSB} formula.

Space-time is knit together by combining an infinite congruence of such time-like trajectories.  For each pair of trajectories, at each time, one must specify an {\it overlap Hilbert space}, ${\cal O}$, whose dimension determines the holoscreen area of the largest causal diamond in the intersection of the diamonds of the individual observers.  ${\cal O}$ is a (usually) proper tensor factor of each of the individual diamond Hilbert spaces.   The fundamental dynamical constraint of HST is that the dynamics and initial conditions in each trajectory Hilbert space must be chosen in such a way that the two density matrices in ${\cal O}$ specified by the individual trajectory time evolution operators and initial conditions, are unitarily equivalent for every overlap.  The pattern of overlaps and their dimensions, completely specifies the causal structure and conformal factor of a Lorentzian geometry.

For space-times with four non-compact dimensions the fundamental variables in a causal diamond of area $\sim N^2$ are complex $N \times N + 1$ matrices $\psi_i^A (P)$.  They should be thought of as spinor spherical harmonics on the two-sphere, with $N$ an eigenvalue cutoff on the Dirac operator on the sphere.  Similarly, the label $P$ labels a set of spinor harmonics on an internal manifold, with an independent cutoff on that Dirac operator.  The anti-commutation relations of the variables are represented on a Hilbert space whose dimension is the exponential of the number of $\psi_i^A$, and this requirement implements the BHFSB relation between area and entropy.

In the limit $N\rightarrow\infty$ one can show\cite{susy} that the commutation relations contain a subalgebra identical to that of the supersymmetry generators on a Fock space of massless superparticles. The Fock space arises, as is familiar from Matrix Theory\cite{bfss} and other large N matrix models from block diagonal $N\times N$ matrices $\psi_i^A \psi_{ A}^{\dagger\ j} $, with blocks of size $K_i$ that go to infinity at fixed ratio.  When $N$ is finite, this leads to an interesting constraint on the concept of particles.   In order to have an approximate Fock space, we need lots of blocks, but in order to localize the states on the sphere, in order to give them a particle interpretation, we need large blocks.  The maximal entropy compromise is reached for $K_i \sim \sqrt{N}$, which gives an entropy $N^{3/2}$.  This scaling of particle entropy is well known from black hole physics.  It represents the maximal entropy in particles within a causal diamond, which will ${\it not}$ collapse to form a black hole.  The rest of the DOF do not have a particle interpretation.  We call them {\it horizon DOF}, and they contribute the overwhelming majority of the entropy of the system.

In a paper in preparation\cite{holounruh} we will show that the horizon DOF are responsible for the Unruh effect, and that integrating them out leads to the S matrix for particle states.   A rough sketch of the argument follows:

For a uniformly accelerated trajectory with acceleration $a$, in a causal diamond in Minkowski space of area $\sim N^2$ we postulate a Hamiltonian

$$H(a) = Z(a) P_0 + \frac{1}{N^2} {\rm tr\ } f(\psi \psi^{\dagger} ) ,$$ where $f(x)$ is a polynomial whose order stays finite as $N \rightarrow \infty$.  $P_0$ converges to the multiparticle energy operator defined by the SUSY commutation relations\cite{susy}.  $Z(0) = 1$ and $Z$ decreases as $a$ increases.  It is of order $1/N$ for the trajectory which skims the ``stretched" horizon.   Standard large $N$ scaling implies that the second term in the Hamiltonian has a bound of order $1/N$.  We assume that the full Hamiltonian, is a fast scrambler, in the language of \cite{sekinosusskind}.  However, in the HST context, in Minkowski space, this is a time dependent Hamiltonian, for the last step in an evolution over a proper time of order $N$.

We define the particle states by a constraint $$\psi_i^A (P) | particle \rangle = 0$$ for of order $K N $ matrix elements, with $ 1 \ll K \ll N$ .   The bilinears formed from the non-zero matrix elements form a block of size $N - K \times N - K$ and blocks of size $K_i$ with $\sum K_i = K $.   The second term in the Hamiltonian preserves the constraint only in the sense of large $N$.   The value of $K$ cannot be changed because the Hamiltonian contains a polynomial of fixed order as $N \rightarrow\infty$ , but the number of nonzero matrix elements can change by an amount of order $K^2$.  This allows the number of blocks to change.  The Hamiltonian $P_0$ is constructed from the individual blocks as in \cite{susy}, and is equal to $\sum K_i$, which is the asymptotically conserved quantum number $K$: the Poincare energy.  For the present paper, the important part about this description of particles is that particle states have, by definition, an {\it entropy deficit} $NK$.  Standard counting arguments, reviewed in \cite{nightmare} show that this is at most $N^{3/2}$.  This is true for particles in empty flat space-time, as well as in the interior of a black hole.  In the latter case, the particle description is only tenable for a time of order the Schwarzschild radius.  

It's worth recalling that the fact that localized states, like particles, correspond to an entropy deficit, first showed up in the holographic description of de Sitter (dS) space.  Indeed, in dS space the entropy formula for a black hole localized near the origin of a causal patch, shows that it decreases the entropy.  The dS vacuum is the maximal entropy state of the system, and any localized excitation has an entropy deficit give by $2\pi R E$ where $R$ is the dS radius and $E$ the energy measured by a geodesic observer. This accounts for the dS temperature.

Returning to our accelerated observers in Minkowski space, recall that $N$ is also the proper time in the causal diamond.  $H(a)$ is actually the time dependent Hamiltonian which generates the last step in time evolution at the beginning and the end of the time symmetric causal diamond.   For the geodesic observer, with $Z = 1$, $H(a)$ generates primarily free motion of the particles.  Note that the horizon states have $P_0 = 0$.   However, in times of order $N$, a correction of order $1/N$ will have a non-trivial effect.  In \cite{holounruh} we will argue that the resulting infinite time S-matrix is unitary on the space of particle states, and that the horizon states give rise to effective interactions between particles, but decouple as $N\rightarrow\infty$, in a way reminiscent of the open strings between D0 branes in \cite{bfss}.   We will use this picture of particle states and interactions in our discussion below.

In passing we note that, as $a$ increases, the coupling between particle and horizon states increases.  This leads to a density matrix for accelerated particles, with a temperature that increases with acceleration; the Unruh effect.

There are three things about the HST description, which differ dramatically from the vague ideas about the way local field theory emerges from QG, that are prevalent in the literature.  One difference is the description of different local observers by different Hilbert spaces, related only through density matrix constraints for shared information. The other two are the related facts that the DOF in the diamond are mostly horizon, rather than particle, states, and that these horizon states have low energy as measured by the Hamiltonian of a geodesic observer. Field theory emerges in a large causal diamond because the particle degrees of freedom can describe many localized excitations on the holographic screen and because they decouple from the horizon states in the large diamond limit.

In string theory, and AdS/CFT, effective quantum field theory (QUEFT) is only valid for computing {\it certain} boundary correlation functions in a limited kinematic range, when the AdS radius is much larger than the string scale.  This has been widely taken as license to use QUEFT whenever we have a space-time region where spatial curvatures and particle energies are small compared to the Planck scale.  In HST this license is revoked.  In a finite causal diamond, there are drastic restrictions on the notion of particle.  The number of particles is bounded in a way that depends on the quantum numbers which will become their energies, and these quantum numbers also restrict the angular localizability of particles.  Moreover, particle states are only a tiny minority of the states in the causal diamond, and the majority, called horizon states, are not at high energy.  Rather, they decouple, approximately, because of the special form of the Hamiltonian sketched above.  Particle states are constrained by an initial condition which sets many horizon DOF to zero.  For geodesic motion, the Hamiltonian preserves this constraint.   QUEFT is only valid for computing transition amplitudes in which both the initial and final states obey the constraints that allow them to be described in terms of particles.  In particular, as we will see, it does {\it not} provide a good model of how black hole states transmute into asymptotic particles as a black hole evaporates.

Before proceeding to discuss an HST model for old black holes, we must address a question of principle.  In HST, the structure of space-time is encoded in non-fluctuating properties of a collection of quantum systems - the dimensions of the overlaps.  A black hole of finite entropy can be formed by a scattering process in Minkowski space.
In HST, the description of this process never, in principle, introduces a space-time structure different from Minkowski space.  The black hole is a quantum state in Holographic Minkowski space, rather than a different space-time geometry.  

We believe, as do the authors of \cite{amps} and \cite{ampsrefs}, that the semi-classical description of particle physics in a dynamical space-time must be a good approximation (in some sense) to the correct quantum system.   What we argue in the next section is that, if we apply the HST description to a classical space-time corresponding to an evaporating black hole, paying careful attention to the way in which QFT emerges from HST, then the firewall phenomenon does not arise.  The models sketched in the following section should thus be thought of as an approximate, effective description of evaporating black holes as a semi-classical geometry, using the HST description of that geometry rather than that of QFT.  We will see that there is no contradiction between the decoupling of particle states from horizon states in causal diamonds where the space-time curvature is low, and the consistency conditions between causal diamonds of observers that fall in to the black hole at very different times.

\section{Are there firewalls in HST?}

We consider a black hole of entropy $N_E^2$ which forms and evaporates in asymptotically flat space-time. We will focus on three detectors, by which we mean time-like trajectories in the resulting space-time.  For a finite entropy black hole, the notion of a classical space-time different from Minkowski space is always an emergent, approximate concept.  We will use it where appropriate, but will find that it must be abandoned near the singularity.  HST gives a completely finite quantum description of the singular regime.

Detector E(arly) falls into the black hole at a time soon after it forms, when the horizon size is of order $N_E$ .  Detector L(ate) falls into the hole some time later, perhaps after the Page time, $N_E^3$.  The horizon size is then $N_L  <  N_E$.  For times before it falls into the hole, we assume that detector L followed a stable geodesic orbit around the black hole, and then ``fired a rocket engine", which accelerated it into the hole. We will follow the time evolution of both detectors for times of order the Page time of the original black hole.   At this point we already encounter a breakdown of classical geometry for detector E.  That trajectory has a proper time, which ``ends" when it is of order $N_E$.  We will assume that the quantum mechanics of this detector continues over the period $N_E$ to $N_E^3$, having a time dependent Hamiltonian, which varies on the Planck time scale.

The causal diamonds of both of these observers are huge, since the trajectories have been around since the infinitely remote beginning of time.  To regularize this infinity, we start from some finite past time $- N$, with $N \gg N_E$.   The Hilbert spaces of both E and L have entropy of order $N^2$.  However, in HST we can specify that tensor factor of the Hilbert space of E or L, which corresponds to the causal diamond from the time the trajectory crossed the horizon, until some given proper time $T$.  The situation is summarized in the diagrams of Figures 1 and 2 .  These figures only cover the relevant part of the Penrose diagram of the evaporating black hole space-time.

\begin{figure}[H]
\center
\includegraphics[width= \textwidth]{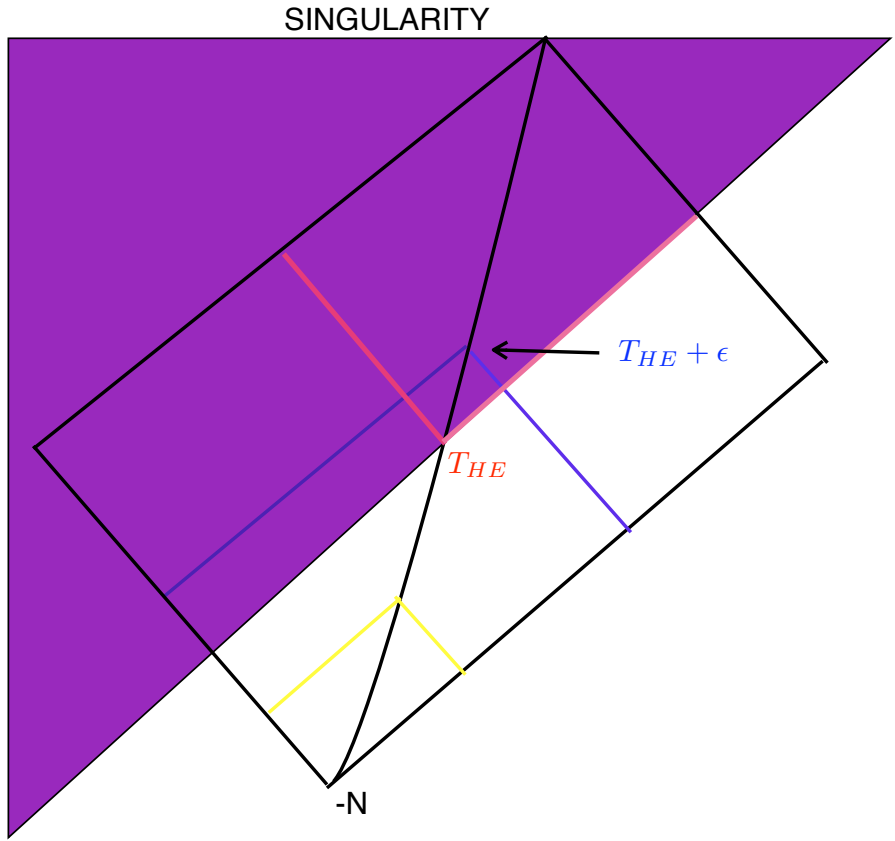}
\caption{ The trajectory E, which falls into a black hole shortly after its horizon forms. Causal diamonds are shown along this trajectory for certain of the times mentioned in the text.  When $\epsilon \sim N_E$ the trajectory hits the singularity. The diamond described by the Hilbert space ${\cal H}_{E\ in} (T_{HE} + \epsilon )$ lies inside the purple black hole interior and its past boundary is indicated in red.}
\label{Trajectory Entering a Newborn Black Hole}
\end{figure}

\begin{figure}[H]
\center
\includegraphics[width= \textwidth]{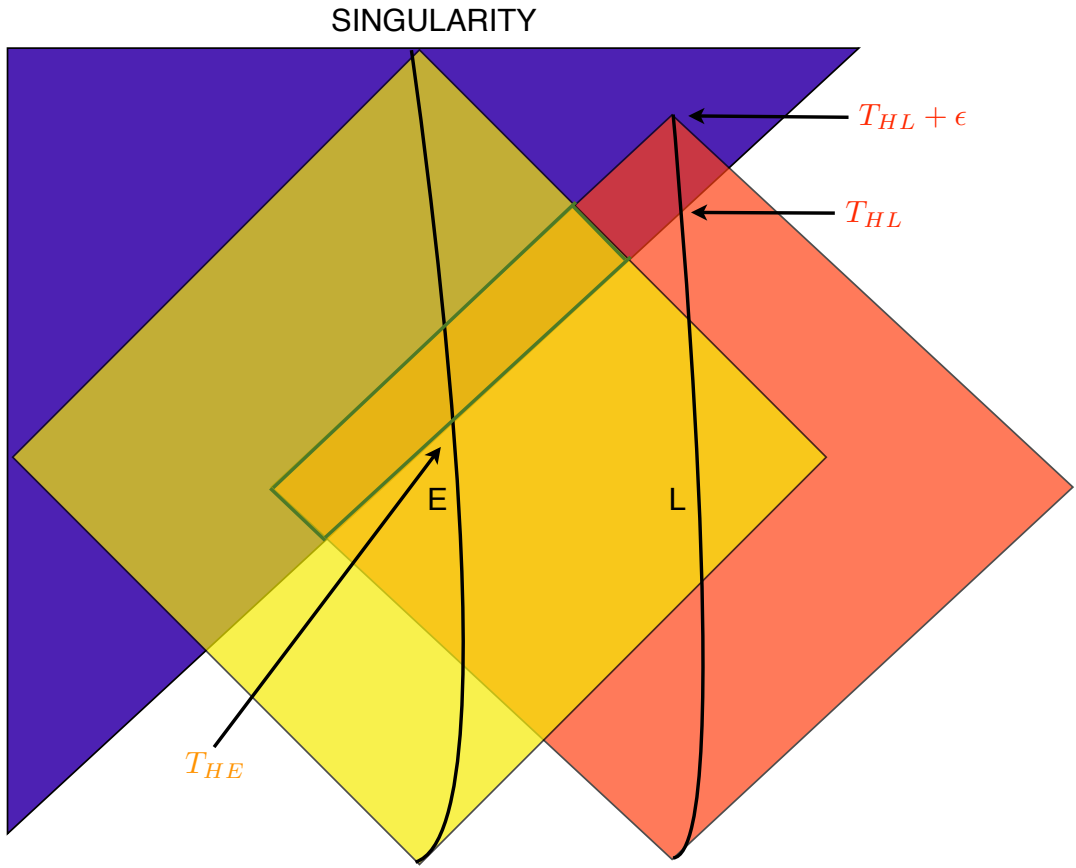}
\caption{This diagram shows both E and L and their overlap. Since the time spanned is of order the Page time, the radius of the horizon (boundary of the blue region) changes between the times at which the two trajectories cross it. The overlap ${\cal O}$ describes the region outlined in green.}
\label{Overlap of the Observers E and L}
\end{figure}
Our final detector, S, will be one which never falls into the black hole.  As we'll see, its description of reality is quite similar to that of L, for a time of order the Page time.  Detector S lives forever, and its asymptotic time evolution operator is the scattering matrix, which is the only gauge invariant observable in the HST theory of Minkowski space.

Let us first examine the dynamics of detector E, up till times of order $N_E$ after the trajectory crosses the horizon.  For times before horizon crossing, the Hamiltonian is very close to that we described for geodesic observers in Minkowski space in the previous section.  It has the form (for $T + N \gg 1)$
$$H_E (T) = P_0 + H_{in}(T) + H_{hor} + (\frac{1}{T + N})^2 V ,$$ where $V$ is the trace of a low order polynomial in the matrix variables.  The Hamiltonian $H_{in}$ acts on a tensor factor of the full Hilbert space of the trajectory E, which consists of some of those states which are not yet in causal contact with the trajectory at the time $T$, namely states that can be described as particles propagating in the interior of the black hole.  The full Hilbert space of the black hole has an entropy $N_E^2$ but that is divided between horizon states, interior particles, and off diagonal variables which are set to zero in the initial state.   As in Minkowski space, the last term in the Hamiltonian excites these states, but they decouple for large $T + N$, leaving over interactions between particles, and between external particles and the black hole.  Before horizon crossing, $H_{in} (T)$ contains only particle energies.  $H_{hor}$ acts on a Hilbert space of entropy $N_E^2 - N_E^{3/2} $ and is a fast scrambler of all those matrix DOF.

After horizon crossing, which occurs at time $T_{HE}$, $H_E (T)$ changes. Classical geometry suggests that $H_{in} (T)$ becomes ``singular" for $T = T_{HE} + N_E$.  We will interpret this as follows.  We've noted that since $N_E \gg 1$, we can separate the variables in the Hilbert space on which $[H_{in} + H_{hor} ](T_{HE} + \epsilon)$ acts into particle and horizon degrees of freedom, and this Hamiltonian will be approximately the same as that of a geodesic observer in a causal diamond of radius $N_E$ in flat space. However, when $\epsilon \sim N_E$ we assume instead that the Hamiltonian has a time dependence on scales of order the Planck time and that the Lie algebra generated by $H_{E\ in} (s)$ for different $s$ is the entire algebra of matrices in a Hilbert space of entropy $N_E^2$.  This implies that the state of the system sweeps out the entire Hilbert space, and that the time averaged density matrix over time scales of order the Planck time is maximally uncertain\cite{bfsing}.   Until $s$ is of order $T_{EH} + N_E$ we can think of it as proper time along the trajectory, but after this there is no geometrical interpretation of this time parameter\footnote{In \cite{bfsing} we considered the possibility that the Hamiltonian became singular in finite time, so that the time of the in-falling QM ends.  The overlap conditions of HST have, so far, been formulated at a synchronized time for all trajectories.  In such a formulation, we must allow the internal time to flow after the singularity is reached, and so we have replaced the singularity by eternal evolution with a time dependent Hamiltonian whose time dependence is Planck scale.  Classical physics gives us no guidance as to which of these formulations is ``correct".  It might be that we could use the singular time dependence, with a time that ends, by generalizing the rules of HST to refer to overlaps between diamonds that are not synchronized.}.

The rest of the Hamiltonian remains close to that of a geodesic observer in flat space for all time. This is enforced by overlap conditions with the Hilbert spaces of trajectories like S, which never fall into the black hole.  We do not have a full HST description of flat space, so we cannot describe the mathematical details of these conditions. 

We can now use this framework to describe the evolution of the black hole as it evaporates.  Roughly speaking, we must make the Hilbert space on which $H_{in} + H_{hor}$ acts shrink, while the complementary tensor factor grows.  This simply means that the time dependent Hamiltonian couples together different degrees of freedom as time goes on.  At a time of order the Page time, the inside tensor factor has entropy $N_L^2 = c N_E^2$, with $c < 1/2$.   We do not have a detailed model of the evaporation process, which would show graphically how the horizon degrees of freedom which originally described the interior of the black hole after the singularity is reached, transmute into particle degrees of freedom.  However, for our present purposes we will need only simple counting arguments, plus the observation that the horizon states which turn into asymptotic particles do not behave like particle states at all.  The authors of \cite{amps} insist on a QUEFT description of localized states in causal diamonds of size $N_E$, as the black hole evaporates.  They use Hawking's derivation of the thermal spectrum of Hawking radiation to justify the assertion that knowing even a little information about the internal black hole state implies that ``high energy modes localized in the diamond", are not in the Minkowski vacuum state of QUEFT, which implies the existence of a firewall.

HST does not allow us to come to this conclusion.  It's simply untrue, in the HST formalism, that most of the states in the diamond are particle states.   The microscopic mechanism of Hawking radiation is the conversion of horizon states into asymptotic particle states.  This occurs because the time dependent Hamiltonian subjects fewer and fewer of the black hole DOF to the scrambling Hamiltonian, and instead includes them among the DOF whose Hamiltonian decouples particles from the horizon of the large causal diamond.   As time goes to infinity, this decoupling becomes complete and the evolution operator is unitary on the space of asymptotic particles.  By this time, the time dependence of the Hamiltonian has eliminated all of the ``black hole behavior".

We do not think that the fact that Hawking's derivation gives the thermal spectrum of Hawking radiation requires us to accept the microscopic picture of the mechanism for production of thermal radiation given by QUEFT, as correct in a theory of Quantum Gravity.  Black holes are large thermal systems, and any model of them which gets their entropy and energy right, must reproduce the approximately thermal nature of the radiation.  HST leads us to expect that the role of QUEFT in large but finite low curvature causal diamonds is much more constrained than has been previously supposed.  The model of black hole states as field theory states around a stretched horizon gets the entropy right, but there is no sacred principle which tells us that anything else about it is correct.  HST is an alternative approach to black hole physics, which gives a radically different description of the microscopic dynamics, but is designed to get the thermodynamics right.   

Now we turn to the description of the trajectory L which follows a stable geodesic orbit around the black hole until a time of order the Page time, and then accelerates inward to encounter the singularity.  Classical physics would lead us to believe that the description of this observer is, when $N_L \gg 1$ the same as that of the observer E.  In particular, the particle physics seen by this observer should be only mildly perturbed by the black hole, until a time of order $T_{HL} + N_L$, where $T_{HL}$ is the time that L's trajectory crosses the horizon.  Note that for this trajectory, the time parameter in the quantum Hamiltonian {\it can} be identified with the proper time along the trajectory, up until $T_{HL} + N_L$.

The essence of the claim of \cite{amps} is that this classical expectation is wrong.  Their argument relies on general properties of a hypothetical theory of quantum gravity which can describe approximate local observables.  In particular, it claims that the system has a single Hilbert space and that different observers correspond to different, possibly incompatible, quantum observables in that system. As we have seen, HST does not have such a structure.  Different time-like trajectories correspond to different Hilbert spaces, and the formalism prescribes only constraints on density matrices for information shared by different trajectories.

A key argument of \cite{amps}, which is not valid in HST, is the assumption that the black hole was in a generic state before evaporation.  Instead, as we've seen, the contention that particle physics is a valid approximation to physics seen by the detector E, for a time of order the Schwarzschild time $N_E$ implies that, at that initial time, the state was special, with an entropy deficit related to its particle content, but bounded by $N_E^{3/2}$.  Since $N_E^2 \gg N_E^{3/2}$ there is no contradiction between this contention and the claim that, from the outside, the black hole behaves like a thermal system, with entropy $N_E^2$. \footnote{ A simple example of such a phenomena can be seen in ordinary physics:
Consider a pool of lava in a deep underground shaft.   It's cooling by emitting photons and boiling off some hot gas.   The photons get out easily, but some of the hot gas goes up another branch of the shaft .   It has much less entropy than half the entropy of the original pool of hot lava.   Most of the energy gets emitted as photons.   Let's further suppose that the trapped gas molecules were emitted in an early blast of the volcano, before the whole thing had a chance to thermalize so that their quantum state is not entangled with that of most of the hot lava as it starts to cool by emitting photons. Eventually the gas molecules fall back down into the lava pool and mix with it.  We end up with everything inside the volcano in its ground state, all the information transmitted out in the photon wave function.   But some of the late photons came from stuff that only got thermalized very late.   The complete final state of the photons is pure, but there's a period of time when an observer could climb down into the volcano and discover the pocket of hot gas, before it fell down and got entangled with the rest of the lava.   Since the pocket of gas had entropy negligible compared to that of most of the lava, the thermodynamics of this process is indistinguishable from that where we imagine there was no such pocket of gas. } In order to show this, we consider a whole sequence of observers, with $N_L < N_E$, for all times between formation of the black hole and the Page time.

In the Hilbert space of detector L, we have to pay a little more attention to causality constraints.   At times before L's trajectory crosses the horizon of the old black hole, its causal diamond is space-like separated from states inside the black hole.  Thus, our model for this system at times longer than the scrambling time of the young black hole is the following.  The Hilbert space is a tensor product of a Hilbert space ${\cal H}_{in}$ of entropy $N_L^{3/2}$, a space ${\cal H}_{hor}$ of entropy $N_E^2 - N_L^{3/2}$, and a space ${\cal H}_{out}$ of entropy $N^2 - N_E^2$.   $N_L$ is the Schwarzschild radius of the old black hole. The space ${\cal H}_{out}$ describes particle physics outside the black hole, plus a set of almost decoupled horizon DOF, which mediate interactions between particles and between particles and the black hole ({\it e.g.} the Newtonian potential of the black hole).   The Hamiltonian in ${\cal H}_{hor}$ is a fast scrambler.  Until observer L's trajectory hits the horizon, the Hamiltonian that describes its causal diamond cannot couple the DOF in ${\cal H}_{hor} \otimes {\cal H}_{out} $, to those in ${\cal H}_{in}$.  This is consistent with a description of time evolution in ${\cal H}_{in}$ in which particle degrees of freedom remain decoupled from horizon degrees of freedom during the entire period from black hole formation until the trajectory of L crosses the horizon.  

It's important to remember that, in HST, everything that goes on inside ${\cal H}_{in}$ before L crosses the horizon is unphysical from the point of view of detector L.  All measurements performed by L before horizon crossing, involve operators that commute with every operator on ${\cal H}_{in}$, and the requirements of causality, as implemented in HST, tell us that the Hamiltonian of L cannot couple to these operators until horizon crossing.  The only constraints that the formalism might impose on dynamics in ${\cal H}_{in}$ before that time comes from the overlap conditions.  We will discuss these in a moment.

The interactions between ${\cal H}_{hor}$ and ${\cal H}_{out}$ must exhibit the following features, in order to be consistent with the semi-classical description of a black hole, and with unitarity.   The system in ${\cal H}_{hor}$ must be in a generic state after the scrambling time, and before evaporation.   As time goes on, the time dependence of the Hamiltonian effects a transfer of DOF between ${\cal H}_{hor}$ and ${\cal H}_{out}$.   This simply means that the Hamiltonian changes so that fewer matrix element DOF are acted on by the scrambling Hamiltonian.  As this happens, those DOF are transferred to the particle physics Hamiltonian in ${\cal H}_{out}$ .  A full HST description of flat space would be required to describe this process in detail.  We do not have such a description yet, but there is nothing paradoxical, or which violates unitarity, in such a process.   The result of it, after the Page time, would be to fully entangle the state of the DOF remaining in ${\cal H}_{hor}$ with that of the emitted Hawking radiation.  The state in ${\cal H}_{in}$ has not coupled to the rest of the Hilbert space and would only be entangled if one had imposed some acausal {\it a priori}
initial condition.  The dynamical rules of HST do not generate entanglement of this state with the rest of the system.

The entropy of ${\cal H}_{in}$ is much less than that of ${\cal H}_{hor}$ so the tiny entropy deficit will not change the thermodynamic properties of the system, and Hawking's calculation of the black hole temperature and entropy will still be valid.   Now, the point is that nothing in this discussion changes as $N_L^2$ becomes less than $\frac{N_E^2}{2}$.
There is nothing special about the Page time in this model.   Every observer, as long as $N_L \gg 1$ describes the black hole in a slightly non-generic state, in which some of its information is not correlated with the early Hawking radiation.  This was true for early observers and remains true for observers that fall in after the Page time.   The only thing that could force us to abandon this model in HST is a failure of the overlap conditions.

In particular, all of these considerations apply to the detector S, which never falls into the black hole.   The space-time picture of an evaporating black hole implies that there are causal diamonds in the black hole interior, which are space-like separated from the causal diamond of this detector, until the black hole shrinks to Planck size.  Correspondingly, its Hilbert space at any time before that factors into three factors ${\cal H}_{in}\otimes {\cal H}_{hor} \otimes {\cal H}_{out}$.  This detector never comes into contact with states that experience rapid time dependence, but it does experience an interaction with scrambled states on the horizon.  Thus, the Hamiltonian couples only the second two tensor factors,
$$H = H_{hor} + H_{out} + H_{hor/ out} + H_{in}$$. The Hilbert space ${\cal H}_{hor}$ contains slightly less than the total entropy of the black hole, enough less for the remaining tensor factor ${\cal H}_{in}$ to described particles propagating in the interior.  As time goes on, the Hamiltonian makes both of the first two factors shrink (in the sense that the DOF participating in $H_{hor} + H_{hor/ out} + H_{in}$ become less numerous, with a corresponding increase in those participating in $H_{out}$.  The entropy ratio of ${\cal H}_{in}$ to ${\cal H}_{hor}$ scales like $N_L^{-\frac{1}{2}}$. The dynamics in ${\cal H}_{in}$ can be chosen to satisfy overlap conditions with all of the infalling observers.

To analyze the overlap conditions, we first
consider a pair of observers, which do not fall into black holes in an asymptotically flat space-time. The late (infinite) time causal diamonds of two such observers coincide, and they share the totality of information in their Hilbert spaces.
The rules of HST imply that the S matrices of the two observers are identical to each other.  For trajectories which fall into black holes, HST only requires us to discuss the overlap conditions.  Consider the parts of the causal diamonds of E and L, which are inside the horizon, at a very short time $\epsilon$ after L crosses the horizon of the old black hole.  It's clear that as $\epsilon \rightarrow 0$ the overlap between these interior causal diamonds is empty.   In general, for small $\epsilon$ the overlap will have an entropy $c \epsilon N_L$, with $c \sim 1$. This follows from dimensional analysis and the fact that the classical geometry of an evaporating black hole is non-singular at the horizon\footnote{Of course, this contention is precisely what \cite{amps} is contesting.  What we show is that if we assume the geometry is non-singular, that assumption is self consistent in HST.}.  Thus the overlap area is linear in the dimensionfull proper time $\epsilon L_P$.  Since the classical geometry depends only on the Schwarzschild radius, our result follows.

What is the density matrix of the system E on the tensor factor ${\cal O}$ of its Hilbert space, which represents this overlap?
According to Page's argument\cite{page} this density matrix is typically the maximally uncertain density matrix as long as $c \epsilon < 1/2 N_L$.  Note that most of the entropy of both causal diamonds is carried by horizon states, for which genericity assumptions do make sense. Since the dynamical evolution of E is exploring its entire Hilbert space, on the Planck time scale, there will be occasional ``glitches" where the overlap Hilbert space is less entangled with the remainder of E's space.  However, these glitches occur on a recurrence time scaling like the exponential of the entropy difference between ${\cal O}$ and its tensor complement.  Thus even though the natural time scale for exploration of the full Hilbert space is the Planck time, these recurrence times will be much longer than $\epsilon$.

In order to make the description of $L$ compatible with classical expectations, we have to be sure that the Hamiltonian
$$H_L (\tau ) = P_0 + \frac{1}{\tau^2} V ,$$ is a good description of the physics of $L$.  Here $\tau = N + T_{page} + \epsilon$, and $T_{page} \sim N_E^3$ is the Page time of the original black hole.   The state of the system L must not involve large correlations between the $o(N_L^{3/2} )$ particle degrees of freedom, and the remainder of the degrees of freedom, which we associate with the horizon.  The latter are in a generic and generically time dependent state.   It is then a simple counting exercise to see that the density matrix on ${\cal O}$ will, with overwhelming probability, be maximally uncertain.

It is important to note that, as long as $\epsilon$ is small, we are completely free to {\it define} ${\cal O}$ as a sub-tensor factor of the horizon states in the full Hilbert space of the system L.  Particle states are defined by the special form of the Hamiltonian, and by the requirement that they have only small correlations with the horizon states.  The rules of HST say that any choice of ${\cal O}$ compatible with the dynamics, which satisfies the overlap conditions, is acceptable.   We learn from this that the overlap between the Hilbert spaces of E and L in an old black hole, must consist entirely of states on L's horizon, which have no effective field theory interpretation.   Classical intuition does not apply to these states.

This situation persists until $c \epsilon N_L = \frac{1}{2} N_L^2$.  After this time the density matrix on ${\cal O}$ becomes time dependent, with Planck scale time dependence. We can say that the system L has hit a firewall, but at a time scale of order the Schwarzschild radius of the old black hole.
Note however that it is quite plausible that the firewall first affects the horizon degrees of freedom.   That is, after $\epsilon = \frac{1}{2c} N_L$, we must modify the Hamiltonian for $k N_L^2$ of the DOF of L, in order to satisfy the overlap conditions.  Here $\frac{1}{2} < k < 1$, and because of the second inequality we are still free to exempt the $N_L^{3/2}$ particle DOF from this modification.  As indicated in the previous section, we believe that, in asymptotically flat space, particle interactions arise from ``integrating out" the horizon DOF.  Thus, the replacement of the Hamiltonian of more and more horizon DOF by one with rapid time dependence, should modify the effective Lagrangian for particles near the singularity.  This is exactly what we expect from the predictions of effective field theory.  Near the singularity, irrelevant operators become important.

Finally, we must discuss the treatment of regions far from the black hole in HST.   In the discussion of \cite{amps} it is the comparison of the entanglement between the observers L and E, with that between L and a distant observer, which leads to a paradox from which one infers the existence of a firewall.    We want to study the system at the time we called $\tau$, which for L is a time of order $\epsilon $ after L's trajectory crosses the horizon of the old black hole.  In HST the overlap constraints for a distant observer and E (or L) obviously refer to the part of the Hilbert space of the observer E, which we called ${\cal H}_{E\ out}$. There is an analogous tensor factor ${\cal H}_{L\ out}$, for the observer L.   The overlap condition for the distant observer S, and L, is simply the statement that the density matrix that L ascribes to ${\cal H}_{L\ out}$ is unitarily equivalent to that assigned by the distant observer to the isomorphic tensor factor in its own Hilbert space.   

We do not yet have a full formulation of HST for particle interactions in asymptotically flat space.  However, since the semi-classical description of Hawking radiation never leads to bizarre behavior far from the black hole, the density matrices assigned to each observer for this Hilbert space should be essentially the thermal density matrix of Hawking particles.   From the point of view of the observer L, these Hawking particles at time $\tau$ are entangled with the interior states of the black hole, in L's description.  As we have seen, as long as $\epsilon < \frac{1}{2c} N_L$, the constraint of compatibility with the description of the same information by E does not require L to encounter any kind of singularity or high energy behavior.   

The compatibility conditions between E and the distant observer refer to a different density matrix for a different tensor factor of the distant observer's Hilbert space.  That tensor factor describes the maximal causal diamond in the intersection of the entire causal diamond of E, and that of the distant observer.  This diamond is entangled, by E's dynamics, with the system inside the horizon.  In E's description, the system inside the horizon has ``long ago hit the singularity".  That is to say, it is subject to a rapidly time dependent Hamiltonian, which mixes all the states in its Hilbert space (which at time $\tau$ has entropy $N_L^2$).  The density matrix of this interior system, averaged over a few Planck times, is maximally uncertain.  This system is coupled to the states in ${\cal H}_{E\ out} (\tau )$ by the Hamiltonian $H_{E\ int} (\tau )$.   Again, we do not have a fully consistent HST description of particles in flat space, but general rules of thermal physics would suggest that it just produces a thermal state.   The only thing that is slightly mysterious is the value of the Hawking temperature.  At first sight one might have imagined that this sort of rapidly time dependent Hamiltonian would lead to a very high temperature.  This kind of dimensional analysis is misleading for ordinary laboratory systems.  The Hamiltonians of atoms have eV scale energy splittings, but the interaction of bulk systems made of atoms with their environment can produce much lower temperatures.   It is really the Hamiltonian $H_{E\ int} (\tau )$, which controls the temperature to which the system in ${\cal H}_{E\ in}$ heats ${\cal H}_{E\ out}$.

We can think of this as an application of the Born Oppenheimer approximation.  The rapid time dependence of the Hamiltonian $H_{E\ in} (\tau)$ is washed out if $H_{E\ int} (\tau )$ couples predominantly to low energy DOF in ${\cal H}_{E\ out}$.   These are the low energy Hawking particles and there is no apparent problem in constructing a model in which they will have the same density matrix as is assigned to them by the distant observer.

In summary, by paying attention to the fact that different observers live in different Hilbert spaces, connected only by relations between density matrices for shared information, and that the observers E and L do not share much information until $\epsilon$ is of order $N_L$, we have shown that there is no necessity to invoke a firewall in HST.  A crucial part of our argument was the fact that in HST, any state describable by particle physics is a state of low entropy, but that the entropy deficit is not enough to affect thermodynamics.  It does however, contradict AMPS' assumption that the remnant black hole is in a generic state after the Page time.  
The most interesting thing about the arguments of \cite{amps} is that they show that a hypothetical theory of quantum gravity which exhibits both locality and unitarity in a single Hilbert space are incompatible with the expectations of classical physics for large but old black holes.   Unitary single Hilbert space theories refer only to the observations of observers that live forever in a space-time with asymptotically frozen boundaries.   The more general form of quantum gravity is that given by HST.  It involves many Hilbert spaces and partially shared information, but appears to be completely consistent with classical physics.  The arguments of AMPS, by which we have been more convinced by extensive discussions with two of the authors, seem instead to lead to a gross violation of classical intuition.  We leave it up to the reader to decide which model he or she prefers.
\vskip.3in
\begin{center}
{\bf Acknowledgments }
\end{center}
T.B. would like to acknowledge conversations with J. Polchinski, L.Susskind and D.Harlow about firewalls.  The authors also acknowledge important email exchanges with D. Marolf and J. Polchinski, which led to considerable clarification of our ideas. The work of T.B. was supported in part by the Department of Energy.   The work of W.F. was supported in part by the TCC and by the NSF under Grant PHY-0969020

\end{document}